\documentclass[%
 reprint,
 amsmath,amssymb,
 aps,
]{revtex4-2}
\usepackage[T1]{fontenc}
\usepackage{graphicx}% Include figure files
\usepackage{comment}
\usepackage{xcolor}
\usepackage{dcolumn}% Align table columns on decimal point
\usepackage{bm}% bold math
\usepackage{hyperref}%
\usepackage{amsmath} 
\usepackage{amssymb}
\usepackage{amsfonts}
\usepackage{mathrsfs}
\usepackage{braket}
\usepackage{float}
\usepackage{multirow}
\usepackage{subfig}
\newcommand{\db}{2\nu\beta\beta}
\newcommand{\zb}{0\nu\beta\beta}
\newcommand{\nm}{\mathcal{M}^{2\nu}}
\newcommand{\zm}{\mathcal{M}^{0\nu}}

\newcommand{\ger}{{}^{76}\text{Ge}}
\newcommand{\as}{{}^{76}\text{As}}
\newcommand{\se}{{}^{76}\text{Se}}

\newcommand{\projang}{\hat{P}^{I}_{MK}}

\newcommand{\be}{\begin{equation}}
\newcommand{\ee}{\end{equation}}
\newcommand{\ba}{\begin{array}}
\newcommand{\ea}{\end{array}}
\newcommand{\bn}{\begin{eqnarray}}
\newcommand{\en}{\end{eqnarray}}

\begin{document}

\preprint{APS/123-QED}
%\title{Density-functional-theory estimate of the $2\nu\beta\beta$ nuclear matrix element in $^{76}$Ge}
\title{Skyrme SV density-functional analysis of the $2\nu\beta\beta$ decay in $^{76}$Ge}

\author{Jan Miśkiewicz}
\affiliation{Institute of Theoretical Physics, Faculty of Physics, University of Warsaw, ul. Pasteura 5, PL-02-093 Warsaw, Poland}
\email{j.miskiewicz@uw.edu.pl}
% \affiliation{Interdisciplinary Centre for Mathematical and Computational Modelling, University of Warsaw}

\author{Jakub Wysocki}
\affiliation{Institute of Theoretical Physics, Faculty of Physics, University of Warsaw, ul. Pasteura 5, PL-02-093 Warsaw, Poland}
 
\author{Wojciech Satuła}
\affiliation{Institute of Theoretical Physics, Faculty of Physics, University of Warsaw, ul. Pasteura 5, PL-02-093 Warsaw, Poland}

%\date{\today}

\begin{abstract}
We present a theoretical study of the two-neutrino $0^+ \rightarrow 0^+$ double beta decay of $^{76}$Ge within the No-Core Configuration-Interaction framework based on the Skyrme SV density functional. We analyze three allowed decay scenarios distinguished by the $[n,m] \equiv [(\nu g_{9/2})^n, (\pi g_{9/2})^m]$ occupancy of the $0g_{9/2}$ intruder orbital, which remains conserved to high precision, as well as by the triaxiality of the daughter nucleus.
The resulting $2\nu\beta\beta$ nuclear matrix element is found to depend strongly on the scenario. For the energetically favored $[4,2]$ occupancy, we obtain $|\mathcal{M}^{2\nu}| = 0.069(7)$~MeV$^{-1}$. For the $[6,0]$ occupancy, the matrix element further depends on the triaxiality parameter $\gamma$ of the two coexisting, closely lying minima in $^{76}$Se, yielding $|\mathcal{M}^{2\nu}| = 0.040(4)$~MeV$^{-1}$ at $\gamma = 17.7^\circ$ and $|\mathcal{M}^{2\nu}| = 0.22(2)$~MeV$^{-1}$ at $\gamma = 41.9^\circ$.
The latter result is consistent with the empirical value reported by A. S. Barabash, $|\mathcal{M}^{2\nu}| = 0.204(14)$~MeV$^{-1}$, while the two former results are comparable to existing calculations based on energy-density-functional frameworks.
Our calculations reveal challenges in the precise determination of the $|\mathcal{M}^{2\nu}|$  for the $^{76}$Ge decay. The structural complexity, triaxiality, and shape coexistence identified in the analyzed nuclei imply a strong sensitivity to fine details of the interaction and configuration mixing. This, in turn, explains the difficulties in theoretical modeling of the $|\mathcal{M}^{2\nu}|$ matrix elements for the $^{76}$Ge decay, which vary by almost an order of magnitude in the available literature.
\end{abstract}

\maketitle

%\tableofcontents

\section{\label{intro}Introduction}

Neutrinoless double beta decay (0$\nu\beta\beta$) is a hypothetical nuclear process in which two electrons are emitted without accompanying antineutrinos.  Its observation would imply that neutrinos are Majorana fermions, i.e., identical to their antiparticles, and would constitute a violation of lepton number conservation. In turn, it would provide direct access to the absolute neutrino mass scale and shed light on leptogenesis as a possible mechanism for generating the matter–antimatter asymmetry of the Universe.

The potential of $\zb$ to probe physics beyond the Standard Model has motivated an extensive experimental program targeting this, so far elusive, extremely rare decay. In particular, 
searches using $^{76}$Ge have achieved leading sensitivity \cite{(DAn21)}. The GERDA and MAJORANA Demonstrator experiments established the feasibility of operating enriched  germanium detectors with ultra-low backgrounds and high energy resolution, setting lower limits on the half-life exceeding $10^{26}$ yr. The next-generation 
experiment LEGEND~\cite{(Bur25)}, which integrates the technologies developed by GERDA \cite{(Ago23b)} and MAJORANA \cite{(Arn23)}, aims to extend the sensitivity to half-lives beyond  $10^{27}$–$10^{28}$ yr, entering the discovery regime for $0\nu\beta\beta$ decay.

Reliable predictions of $\zb$ decay half-lives, essential for guiding experimental searches, require accurate evaluation of nuclear matrix elements (NMEs). These must be computed within nuclear many-body frameworks. However, the current situation remains unsatisfactory, as different approaches often yield results that differ by factors of a few. Identifying the origins of these discrepancies, i.e., the limitations of specific models, is therefore crucial for progress in the field.

It has been suggested that the performance of theoretical models in describing subtle aspects of nuclear structure may be assessed indirectly through the measured two-neutrino double beta decay ($\db$) channel. In this context, a possible correlation between the $\db$ and $\zb$ decay modes has been discussed. In particular, efforts have focused on identifying systematic relationships between $\db$ NMEs and their predicted $\zb$ counterparts \cite{(Rod03a),(Sim11),(Sim13)}. Calculations of $\db$ NMEs may thus provide an important consistency check for nuclear models employed in $\zb$ studies.

The aim of this work is to employ the density-functional-based no-core configuration-interaction (DFT-NCCI) framework~\cite{(Sat16d)} to compute the nuclear matrix element for the $\db$ ($0^+ \rightarrow 0^+$) decay $^{76}$Ge $\rightarrow$ $^{76}$Se. The formalism was recently applied to the $\db$ decay of $^{48}$Ca, where the situation was found to be relatively simple and well controlled~\cite{(Mis25b)}. In that case, the simplicity of the underlying shell structure leads to consistent results across different nuclear models, in agreement with the empirical matrix element of Barabash~\cite{(Bar20)}.

The situation in the $A=76$ region appears to be markedly different. Nuclear matrix elements reported in the literature show substantial spread, in some cases approaching an order of magnitude. This likely reflects the richness of nuclear structure effects in this mass region, including pronounced deformation, triaxiality, shape coexistence and mixing.

The DFT-NCCI framework, with its natural capability to describe deformation and shape coexistence, provides a suitable tool to investigate these effects. In this work, we focus on exploring different structural scenarios to assess their impact on the calculated matrix elements for the $\db$ decay. In this context, it is worth noting that the role of deformation in the suppression or enhancement of neutrinoless NMEs has been highlighted previously, see, e.g., Ref.~\cite{(Men09)} and references therein.

This article is organized as follows.
In Sect.~\ref{theory}, we concisely summarize the theoretical model applied to the analysis of the $\ger$ $\db$ decay. In Sect.~\ref{confs}, we describe two possible decay scenarios corresponding to different classes of virtual transitions, together with shape coexistence in the ground-state configuration of $\se$. Section~\ref{results} discusses the results of the numerical calculations of the $\db$ NMEs for both scenarios. Finally, the summary and concluding remarks are provided in Sect.~\ref{summary}.

\section{Theoretical framework}\label{theory}

The evaluation of the $\db$ ($0^+ \rightarrow 0^+$) nuclear matrix element follows the Fermi golden rule of the 2nd order. Neglecting the Fermi contribution relative to the dominant axial-vector (Gamow-Teller) term, the master formula for the matrix element reads as \cite{(Tom91)}:
\begin{equation}
\label{master}
    \nm = \sum_{m}\frac{ \langle \psi_{f} \vert \hat{\sigma}\hat{\tau}^{-} \vert  \psi_{m} \rangle\langle \psi_{m} \vert \hat{\sigma}\hat{\tau}^{-} \vert \psi_{i} \rangle }{\Delta E_{m} + \frac{1}{2}Q_{\beta\beta}-\Delta M}.
\end{equation}
Here $\psi_{i}$($\psi_{f}$) stands for the  $I^{\pi}=0^+$ states in the parent(daughter) nucleus, and $\psi_{m}$ is the $I^{\pi}=1^+$ 
state in the virtual intermediate nucleus. $Q_{\beta\beta}$ is the $\db$ $Q$-value, $\Delta E_m$ is the excitation energy of the $m$th intermediate 
state relative to the nucleus' ground state, and $\Delta M = M_{m} - M_i$ is the mass difference between the intermediate ($M_m$)
and parent ($M_i$) nuclei. 

A central part of the theoretical treatment is the choice of nuclear framework used to construct nuclear eigenstates with well-defined quantum numbers. For the construction of the configuration space we applied DFT-NCCI framework, developed by our group. The detailed description of the model may be found in the work \cite{(Sat16d)}, here we will only give a brief summary of the theoretical system.

In this method, one starts by constructing the configuration space --- a set of mean-field Slater determinants $\vert \varphi_ i \rangle$ referring to a nuclear ground state and a selected number of its low-lying excitations of the type 1p-1h, 2p-2h, etc. This step is executed in the SR-DFT with an SV-parametrized Skyrme functional. Next, each configuration is being projected first, on well-defined angular-momentum $I,M,K$ numbers:
\begin{equation}
\label{sr}
    \vert \varphi; IMK;T_z \rangle = \hat{P}^{I}_{MK} \vert \varphi \rangle,
\end{equation}
where $T_z$ denotes $z$-component of total isospin $T$ and $\projang$ is the angular-momentum integral projector \cite{(Sat16d)}. The proper symmetry restoration requires additional mixing of the projected state (\ref{sr}) within $K$ quantum number:
\begin{equation}
\label{mrkmix}
    \vert \varphi; IM;T_z \rangle =  \frac{1}{\sqrt{\mathcal{N}_{\varphi;IM;T_z}}}\sum_{K}a^{I}_{K}  \vert \varphi; IMK;T_z \rangle .
\end{equation}
A specific collection of the above linearly independent states --- \textit{natural states}, may be chosen to span a subspace called \textit{collective space}. Finally, the eigenstates and eigenenergies in the DFT-NCCI model may be determined by solving Hill-Wheeler equation in the collective space. As a result of its diagonalization, one obtains normalized DFT-NCCI states:
\begin{equation}
\label{dftst}
    \vert IM; T_z \rangle^{(n)} = \frac{1}{\sqrt{\mathcal{N}^{(n)}_{IM;T_z}}}\sum_{ij}\eta^{(n)}_{ij} \vert \varphi_i; IM; T_z \rangle^{(j)},
\end{equation}
where the sum runs over all $i$-configurations and index $j$, which denotes consecutive $K$-mixed states of the same $I$ within the fixed $i$-configuration.

\begin{figure}[htb]
\centering
\includegraphics[width=0.45\textwidth]{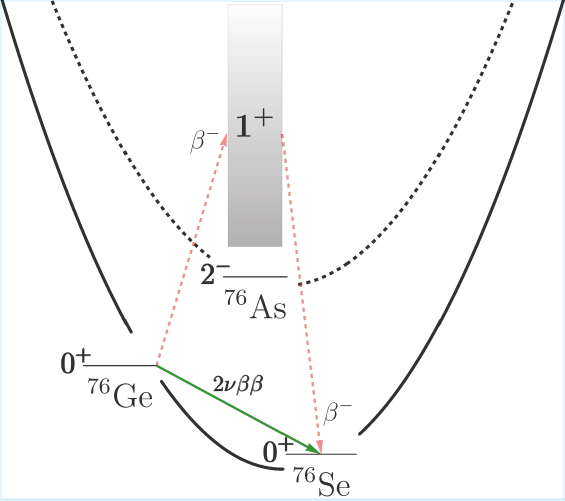}
\caption{(Color online) Depiction of the Fermi golden rule of second order for $A=76$. The virtual configuration space is constructed on consecutive $1^{+}$ excitations of $^{76}$As nucleus.}
\label{fig:Fig01}
\end{figure}

We applied the outlined scheme to compute ground state of the parent $\ger$, daughter $\se$ and the intermediate $I=1^+$ spectrum of $\as$ as depicted on Figure~$\ref{fig:Fig01}$. Translating this into the Fermi golden rule regime, the final formula for the $\nm$ for $\ger \rightarrow \se$ decay reads:
\begin{widetext}
    \begin{eqnarray}
    \label{nm76}
    \nm\big(\ger \rightarrow \se \big) = \sum_{m} \frac{\langle \se; 0^{+}_{f} \vert \sum_{i}\hat{\sigma}_{i}\hat{\tau}^{-}_{i} \vert \as; 1_{m}^{+} \rangle \langle \as; 1_{m}^{+} \vert \sum_{i}\hat{\sigma}_{i}\hat{\tau}^{-}_{i} \vert \ger; 0^{+}_{i} \rangle}{\Delta E_{m} + \frac{1}{2}Q_{\beta\beta}-\Delta M}. 
\end{eqnarray}
\end{widetext}
Here, the index $m$ runs over the $|1^+_m\rangle$ excitations in $\as$; $\Delta E_m$ are the calculated excitation energies of the intermediate 
$|1^+_m\rangle$ states in $\as$ normalized to the lowest experimental $1^+$ state at $\Delta E_1 = 100$\,keV. In the calculation of the matrix element, we use the experimental values for the $Q$-value  $Q_{\beta\beta} = 2039.1$\,keV and for the mass difference $\Delta M = M(^{76}\text{Ge}) - M(^{76}\text{As}) = 410.9$\,keV. The necessary data on mass excess were taken form the NNDC database~\cite{(ensdf_url2)}. 

Within the DFT--NCCI framework, the evaluation of Gamow--Teller (GT) matrix elements is based on expectation values between angular-momentum--projected mean-field Slater determinants, followed by their mixing in the DFT--NCCI basis. For $\beta^-$ decay, it is convenient to express the GT operator as a rank-1 spherical tensor, 
$\hat{\mathcal{O}}^{\mathrm{GT}}_{\mu\nu} = \hat{\tau}^{-}_{1\mu}\,\hat{\sigma}_{1\nu}$.
The matrix elements between $I$-projected Slater determinants $|\varphi; I'M'K'\rangle$ and $|\psi; IMK\rangle$ can then be written as
\begin{equation}
\begin{aligned}
\label{M_GT}
\mathcal{M}^{\mathrm{GT}}_{\mu\nu} = \langle & \varphi; I'M'K'; T_z+1 | \hat{\mathcal{O}}^{\mathrm{GT}}_{\mu\nu} | \psi; IMK; T_z \rangle \\
=\, & C^{I'M'}_{IM,\,1\nu}
\sum_{\xi} C^{I'K'}_{I\,K'-\xi,\,1\xi} \frac{2I+1}{8\pi^2}
\int d\Omega\, D^{I*}_{K'-\xi,\,K}(\Omega) \\
& \times \langle \varphi | \hat{\tau}_{1\mu}\,\hat{\sigma}_{1\xi} | \widetilde{\psi} \rangle,
\end{aligned}
\end{equation}
where $|\widetilde{\psi}\rangle \equiv \hat{R}(\Omega)|\psi\rangle$ denotes the rotated Slater determinant.

The evaluation of $\mathcal{M}^{\mathrm{GT}}$ thus reduces to the computation of the kernels $\langle \varphi | \hat{\tau}_{1\mu} | \widetilde{\psi} \rangle$ and $\langle \varphi | \hat{\tau}_{1\mu}\hat{\sigma}_{1\xi} | \widetilde{\psi} \rangle$ using the Generalized Wick Theorem. However, when only angular-momentum projection is performed, these kernels vanish due to the orthogonality condition
\begin{equation}
\langle \varphi (T_z+1) | \psi (T_z) \rangle \propto \delta_{T_z,T_z+1} = 0.
\end{equation}
This difficulty can be overcome by performing an additional rotation in isospin space of one of the Slater determinants, which aligns the isospin projections and restores nonvanishing overlaps. This procedure introduces an additional integration over the isospin angle $\beta_T$. For details, see Ref.~\cite{(Mis25b)}.

\section{Configuration Space}{\label{confs}}

Within the DFT-NCCI framework, the configuration spaces of the parent, daughter, and virtual intermediate nuclei are built dynamically by adding consecutive excitations constructed upon their ground states until stability of the calculated $\nm$ matrix element is achieved within the adopted uncertainty.

Similarly to Ref.~\cite{(Mis25b)}, the numerical calculations related to the construction of the configuration space, the evaluation of Gamow–Teller matrix elements, and the overlaps between mean-field Slater determinants were performed using a developmental version of the HFODD solver~\cite{(Dob09d),(Sch17),(Dob21)}. As before, a Cartesian basis comprising 12 harmonic-oscillator shells was used, and integration over the Euler angles was performed numerically using 20 knots in each direction.

In contrast to the $A=48$ case, the nuclei participating in the $^{76}$Ge decay are highly triaxial. Hence, $K$ is no longer a good quantum number, and the single-particle states -- and, in turn, the configurations -- are identified by specifying the occupied single-particle levels within the four blocks characterized by different parity-signature quantum numbers $(\pi, r)$~\cite{(Dob09d)}. These correspond to two externally imposed, point-like dichotomic symmetries. Here, $\pi = \pm$ denotes the parity quantum number, while $r = \pm i$ is the eigenvalue of the $y$-signature operator, defined as:
\begin{equation}
    \hat{R}_y = e^{-i\pi\hat{J}_y},
\end{equation}
\begin{equation}
    \hat{R}_y\psi_{i}(\vec{r},\sigma)
    = r_i\psi_{i}(\vec{r},\sigma).
\end{equation}

\subsection{Ground-state configurations in $^{76}$Ge and $^{76}$Se}

The analyzed decay proceeds between the ground states of the parent $\ger$ and daughter $\se$ nuclei. 
It appears that a unique determination of the configurations representing the ground states of $\ger$ and $\se$ 
is not possible due to the limited precision of the SV functional used in our calculations and the intrinsic triaxiality 
of the analyzed nuclei, which renders their structure highly complex.

\begin{figure}[h!]
\vspace{0.2cm}
\centering
\includegraphics[width = 0.44\textwidth]{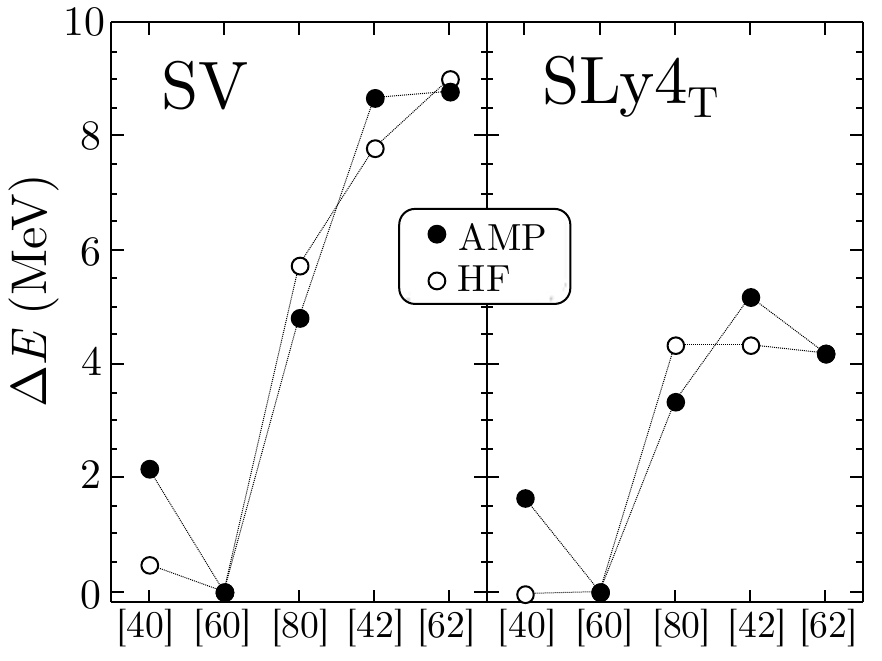}
\caption{\label{fig:Fig02} (Color online)
The lowest-lying mean-field (white dots) and angular-momentum-projected $I=0$ (black dots) solutions 
corresponding to different $[n,m]$ configurations in $^{76}$Ge, calculated using the SV (left panel) and 
$\text{SLy4}_{\text{T}}$ (right panel) Skyrme interactions. 
}
\end{figure}

The configurations -- i.e., the Hartree-Fock (HF) solutions -- in this region can be conveniently classified by specifying the occupancies of the anomalous-parity intruder orbitals: $[(\nu g_{9/2})^n, (\pi g_{9/2})^m]$, or simply $[n,m]$.
In $\ger$, the lowest HF solution, see the left panel of Fig.~\ref{fig:Fig02}, corresponds to the $[6,0]$ configuration.
The $[4,0]$ configuration is excited, but only by approximately 450~keV. The $[6,0]$ configuration is therefore favored by the model; however, the $[4,0]$ configuration cannot be completely disregarded \emph{a priori}, as its excitation energy lies within the typical uncertainties of energy-density-functional-based (EDF) theories.

Note that the other configurations $[8,0]$, $[4,2]$, and $[6,2]$, lie much higher in excitation energy than the $[4,0]$ 
configuration and are therefore less likely. Angular-momentum projection further corroborates the $[6,0]$ assignment, 
which correspond to more deformed nucleus than the $[4,0]$ one. The $I=0$ state projected from the $[4,0]$ 
configuration is 2.1\,MeV higher in energy than its counterpart projected from the $[6,0]$ configuration.

Note that the conclusions drawn above also hold for the SLy4 Skyrme force~\cite{(Cha98)}, as shown in the right panel of Fig.~\ref{fig:Fig02}. This modern parametrization, augmented with tensor terms ($\text{SLy4}_{\text{T}}$), provides a much more reliable description of the configuration assignment than the SV force. At the HF level, the $[4,0]$ and $[6,0]$ configurations are nearly degenerate, with a slight preference for the $[4,0]$ configuration, which nicely illustrates the uncertainties inherent to EDF-based approaches. However, after angular-momentum projection, the $I=0$ state projected from the $[4,0]$ configuration lies 1.6\,MeV above the corresponding state projected from the $[6,0]$ configuration. Note also that the $[4,2]$, $[6,2]$, and $[8,0]$ configurations remain significantly higher in energy and, similarly to the case of the SV force, can therefore be safely ruled out.

\begin{figure}[h!]
\vspace{0.2cm}
\centering
\includegraphics[width = 0.44\textwidth]{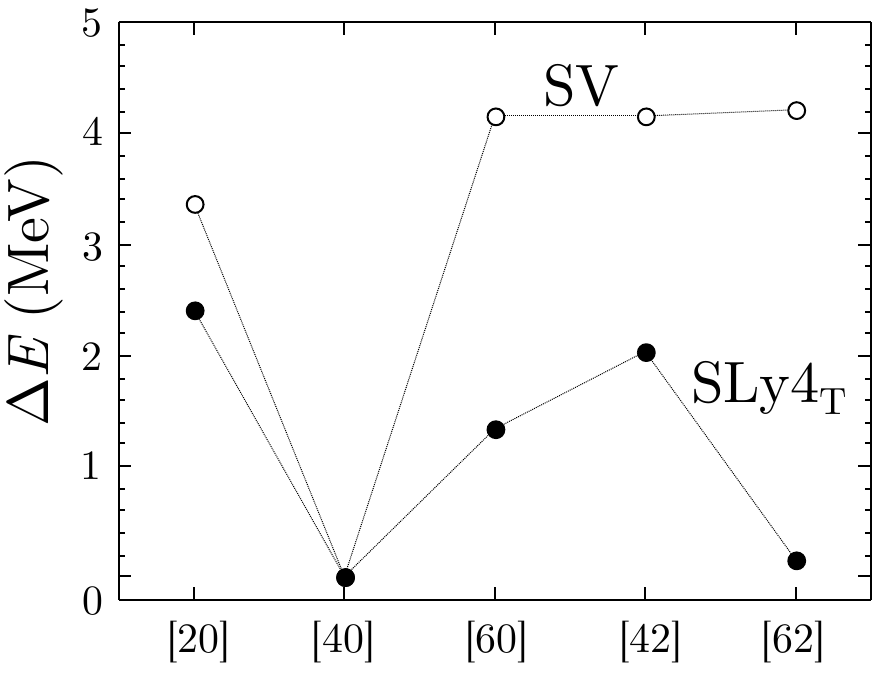}
\caption{\label{fig:Fig03} (Color online)
The angular-momentum-projected $I=0$ states projected from HF solutions corresponding to 
different $[n,m]$ configurations in $^{76}$Se for the SV (white dots) and $\text{SLy4}_{\text{T}}$ (black dots) Skyrme interactions.
}
\end{figure}

The situation in the daughter nucleus $\se$ is even more intricate (see Fig.~\ref{fig:Fig03}). Energetically, the lowest configuration has $[4,0]$ intruder content and corresponds to oblate shape. For the SLy4 force, the $I=0$ states projected from other configurations including $[4,2]$, $[6,0]$, and $[6,2]$ are relatively low in the energy window of order of 2\,MeV as 
seen in Fig.~\ref{fig:Fig03}. Moreover, in case of  $[4,0]$ and $[6,0]$ configurations, shape coexistence is observed, with 
less- and more-triaxial solutions lying relatively close to each other. % This is illustrated in Fig.~\ref{fig:Fig04} for 
%the $[6,0]$ configuration.

For the SV force the scenario is similar though the projected solutions corresponding to excited configurations 
are higher. Again, taking into account typical uncertainties associated with the Skyrme force imply that 
none of the above-mentioned solutions can be excluded \emph{a priori}, except for $[6,2]$, which can only be reached 
via $\db$ decay from the $[6,2]$ or  $[8,0]$ configurations in the parent nucleus - both of which are highly excited, 
as discussed above.

\begin{figure}[h!]
\centering
\includegraphics[width=0.35\textwidth]{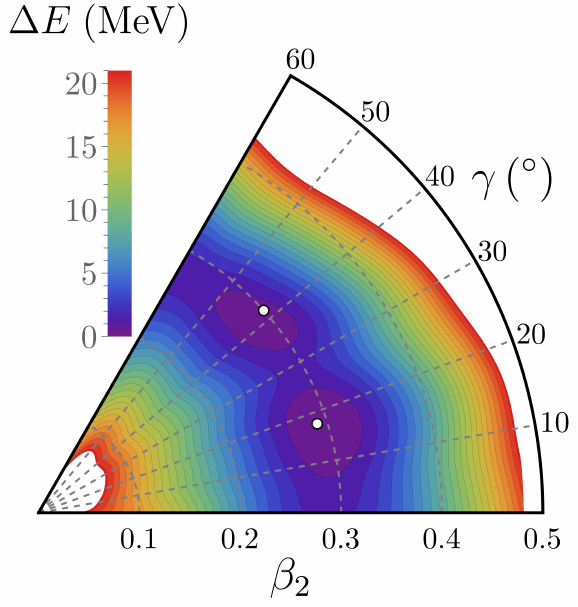}
\caption{(Color online) Shape coexistence in the $\se$ $[6,0]$ configuration shown on the nuclear potential energy surface in the Bohr $(\beta_2,\gamma)$ deformation plane with white markers denoting local minima. The excitation energy has been renormalized to the HF minimum at $\gamma=17.7^{\circ}$.}
\label{fig:Fig04}
\end{figure}

\subsection{The possible decay paths}

The arguments presented above would not hold in the presence of strong mixing between configurations with different intruder contents. Let us note, however, that such mixing would require Cooper-pair scattering from the $g_{9/2}$ intruder orbitals to normal-parity states originating from the $pf$ spherical shells, or vice versa. In principle, this mixing could be effective via a structureless seniority-type pairing interaction. In the present variant of our model, however, the mixing proceeds through the same Skyrme SV interaction that is used to generate the mean-field configurations, and appears to be largely ineffective.

\begin{figure}[h!]
\centering
\includegraphics[width=\columnwidth]{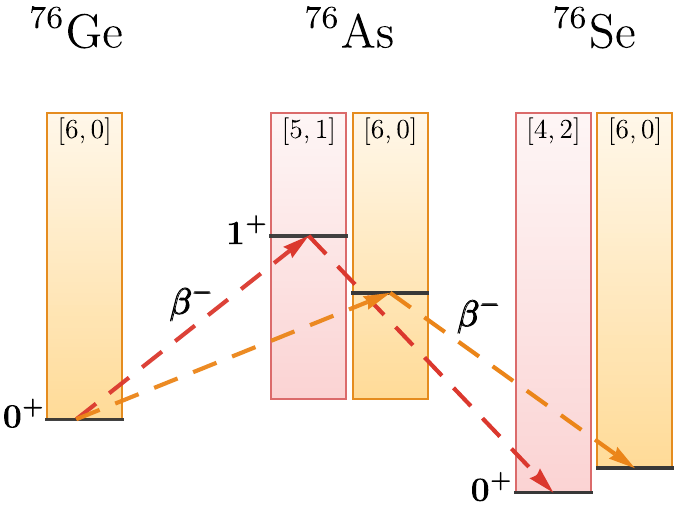}
\caption{(Color online) Schematic illustration of the two $\db$ decay paths analyzed in this work. The dashed arrows denote virtual $\beta$ transitions through the intermediate nucleus. Orange (red) boxes represent configuration spaces built upon excitations that conserve (do not conserve) the $[6,0]$ occupancy of the $g_{9/2}$ orbitals.}
\label{fig:Fig05}
\end{figure}

This implies, that within our model the intruder content can be treated, to a high degree of accuracy, as a conserved quantity. Hence, the mixing can proceed only within the subspace characterized by fixed $[n,m]$ occupation numbers. In turn, within the DFT-NCCI framework used here, the virtual transition can proceed along three disjoint decay paths, denoted hereafter as C40, C42 and C60:
\begin{equation*}
\begin{aligned}
\mathrm{C40}:\ [4,0] &\to [4,0] \to [4,0], \\
\mathrm{C42}:\ [6,0] &\to [5,1] \to [4,2], \\
\mathrm{C60}:\ [6,0] &\to [6,0] \to [6,0].
\end{aligned}
\end{equation*}
In this work, we focus on the C42 and C60 paths (see Fig.~\ref{fig:Fig05}), considering, for the C60 case, the two possibilities corresponding to decay into the less- and more-triaxial solutions illustrated in Fig.~\ref{fig:Fig04}. These two paths originate from the $[6,0]$ configuration in the parent nucleus, which is favored by the model. Let us emphasize that, irrespective of the scenario, the matrix element is evaluated using Eq.~(\ref{nm76}) with the input data given below that formula.

The aim is to explore the sensitivity of the matrix element to the nuclear structure. The analysis of the decay along the 
C40 path will be presented in a forthcoming work, together with the study of the effective coupling constant 
(quenching) in this mass region. Note that the decay along the C42 and C60 paths proceeds through positive- and negative-parity active orbitals, respectively.

\subsection{Configurations in $^{76}$Ge}{\label{76Ge-conf}}

The parent and daughter nuclei are even–even. Hence, we assume that the only configurations that can mix with the ground-state configuration are seniority-zero neutron–neutron ($\alpha_\nu \tilde{\alpha}_\nu)$ and proton–proton ($\alpha_\pi \tilde{\alpha}_\pi$) pairs, where $\alpha_\tau$ and $\tilde{\alpha}_\tau$ denote opposite-signature (time-reversed) single-particle Nilsson states. Such excitations, and the configurations built upon them, carry no alignment.

We also include in the configuration space proton–neutron pair excitations ($\alpha_\nu \tilde{\alpha}_\pi$) and/or ($\tilde{\alpha}_\nu \alpha_\pi$). Due to isospin and axial-symmetry breaking, these configurations carry a non-vanishing, albeit relatively small, alignment. In contrast to the $A=48$ mass region, they appear to have a very weak influence on our results and can therefore be disregarded.

We exclude broken-pair, seniority-two (and higher) configurations from our considerations, as they explicitly violate time-reversal symmetry. Test calculations that include a few of the lowest broken-pair configurations confirm that they hardly mix with the ground state.

Following the general rules outlined above, we include thirty configurations in the configuration space of $^{76}$Ge. It is convenient to divide them into the following groups:
\begin{itemize}
\item
{\bf Group 1} includes the ground-state configuration and the seniority-zero $nn$-pairing $2p$–$2h$ excitations within the Nilsson orbits originating from the $g_{9/2}$ spherical subshell. There are seven basic configurations of this type (including the ground state), all of which are included in the calculations.
\item
{\bf Group 2} consists of five $2p$–$2h$ $nn$-pairing excitations among the Nilsson orbits originating from the $\nu(fp)$ spherical subshells. All configurations of this type are included.
\item
{\bf Group 3} consists of four $2p$–$2h$ $pp$-pairing excitations among the Nilsson orbits originating from the $\pi(fp)$ spherical subshells. There are eight configurations of this type.
\item
{\bf Group 4} contains eleven configurations corresponding to seniority-zero $2p$–$2h$ $nn$-pairing excitations $(\nu g_{9/2})^2 \rightarrow (\nu (sdg))^2$ across the $N = 50$ shell gap.
\item
{\bf Group 5} comprises three configurations. Two of them correspond to the lowest seniority-zero $2p$–$2h$ $pp$-pairing excitations across the $Z = 28$ [$(\pi f_{7/2})^2 \rightarrow (\pi (pf))^2$] and $Z = 50$ [$(\pi g_{9/2})^2 \rightarrow (\pi (sdg))^2$] shell gaps, respectively. The third represents the $np$-pairing configuration.
\end{itemize}

The calculated energy gain in the ground state of $^{76}$Ge due to configuration mixing is very modest, not exceeding 0.250\,MeV.
Moreover, no shape-coexisting low-lying minima within this set of configurations were found.

\subsection{Configurations in $^{76}$As}{\label{76As-conf}}

The configuration space in $^{76}$As consists of 55 $[5,1]$-type configurations relevant for the C42 path. These can be grouped into three 
distinct categories:
\begin{itemize}
\item
{\bf Group 1} consists of seniority $\nu = 2$ configurations obtained by exciting unpaired particles among the Nilsson orbitals originating from the spherical $g_{9/2}$ subshell, as well as a few $4p$–$4h$ $\nu = 2$ configurations involving the excitation of a seniority-zero pair. In total, 39 configurations of this type are included.
\item
{\bf Group 2} includes 8 configurations involving a single neutron excitation across the $N = 50$ shell gap. These are $1p$–$1h$ excitations of the type $(\nu g_{9/2}) \rightarrow (\nu (sdg))$.
\item
{\bf Group 3} comprises 8 configurations involving the lowest cross-$Z = 50$ shell $1p$–$1h$ excitations of the type $(\pi g_{9/2}) \rightarrow (\pi (sdg))$.
\end{itemize}

The general strategy behind the choice of configurations was to include the lowest-lying configurations of each type. In most cases, we explored the potential-energy surfaces for alternative shape-coexisting minima; however, no such minima were found.

Similar strategy was applied to build the configuration space relevant for the C60 path. In this case we calculated 46 configurations, 
which can again be grouped into three distinct categories:
\begin{itemize}
\item
{\bf Group 1} is built predominantly upon seniority $\nu = 2$ in-shell configurations obtained by exciting unpaired particles among the Nilsson orbitals originating from the spherical $pf$ subshells. It also includes a few $4p$–$4h$ $\nu = 2$ configurations involving, in addition, an excitation of a seniority-zero pair.
\item
{\bf Group 2} includes 4 configurations involving the lowest cross-$N = 28$ shell $1p$–$1h$ neutron excitations of the type $(\nu f_{7/2}) \rightarrow (\nu (pf))$.
\item
{\bf Group 3} comprises 6 configurations involving the lowest cross-$Z = 28$ shell $1p$–$1h$ excitations of the type $(\pi f_{7/2}) \rightarrow (\pi (pf))$.
\end{itemize}

Again, the general strategy was to include the lowest-lying configurations of each type, since the energy denominator in Eq.~(\ref{nm76}) suppresses (albeit only moderately) the contribution from high-lying $|1^+\rangle$ states to the $\nm$ matrix element. In contrast to the previous case, we also identified and included 12 shape-coexisting configurations. Cross-shell excitations across $N(Z)=50$ to Nilsson orbitals originating from the $h_{11/2}$ intruder were disregarded.

\subsection{Configurations in $^{76}$Se}{\label{76Se-conf}}

The configuration space in $^{76}$Se for the energetically preferable C42 path includes 42 configurations. 
For the sake of further, detailed discussion of the stability of the $\db$ matrix element the 
configurations, apart of the ground-state, are divided into the following groups:
\begin{itemize}
\item
    {\bf Group 1} includes  the seniority-zero in-shell $nn$-pairing $2p$–$2h$ excitations within the 
Nilsson orbits originating from the $g_{9/2}$ spherical subshell. There are 6 basic configurations of this type (excluding the ground-state), 
all of which are included in the calculations. 
\item
    {\bf Group 2} includes seniority-zero $pp$-pairing $2p$–$2h$ excitations within the 
Nilsson orbits originating from the $g_{9/2}$ spherical subshell. There are 5 configurations of this type and 4 of them are included. 
\item
    {\bf Group 3} comprises 3  lowest $4p-4h$ seniority zero configurations. They have negligible effect of the results.  
\item
    {\bf Group 4} consists 5 (all) $2p$–$2h$ $nn$-pairing excitations among the Nilsson orbits originating from the $\nu(fp)$ spherical 
    subshells. 
\item
    {\bf Group 5} consists 7, out of 9, of $2p$–$2h$ $pp$-pairing excitations among the Nilsson orbits originating from the 
    $\pi(fp)$ spherical subshells. 
\item
    {\bf Group 6} contains  10 configurations corresponding to $np$-pairing excitations.  They essentially do not mix with the ground-state and 
    have no effect of the matrix element.
\item
    {\bf Group 7} comprises 6 configurations. Four of them correspond to the lowest seniority-zero $2p$–$2h$ $nn$-pairing excitations 
    $(\nu g_{9/2})^2 \rightarrow (\nu (sdg))^2$ across the $N = 50$ gap. The space includes also  the lowest $pp$-pairing excitation
    $(\pi f_{7/2})^2 \rightarrow (\pi (pf))^2$ across the $Z=28$ gap and the lowest excitation across the $Z = 50$ shell gap,
    $(\pi g_{9/2})^2 \rightarrow (\pi    (sdg))^2$. 
\end{itemize}

\begin{figure}[h!]
\centering
\includegraphics[width=0.48\textwidth]{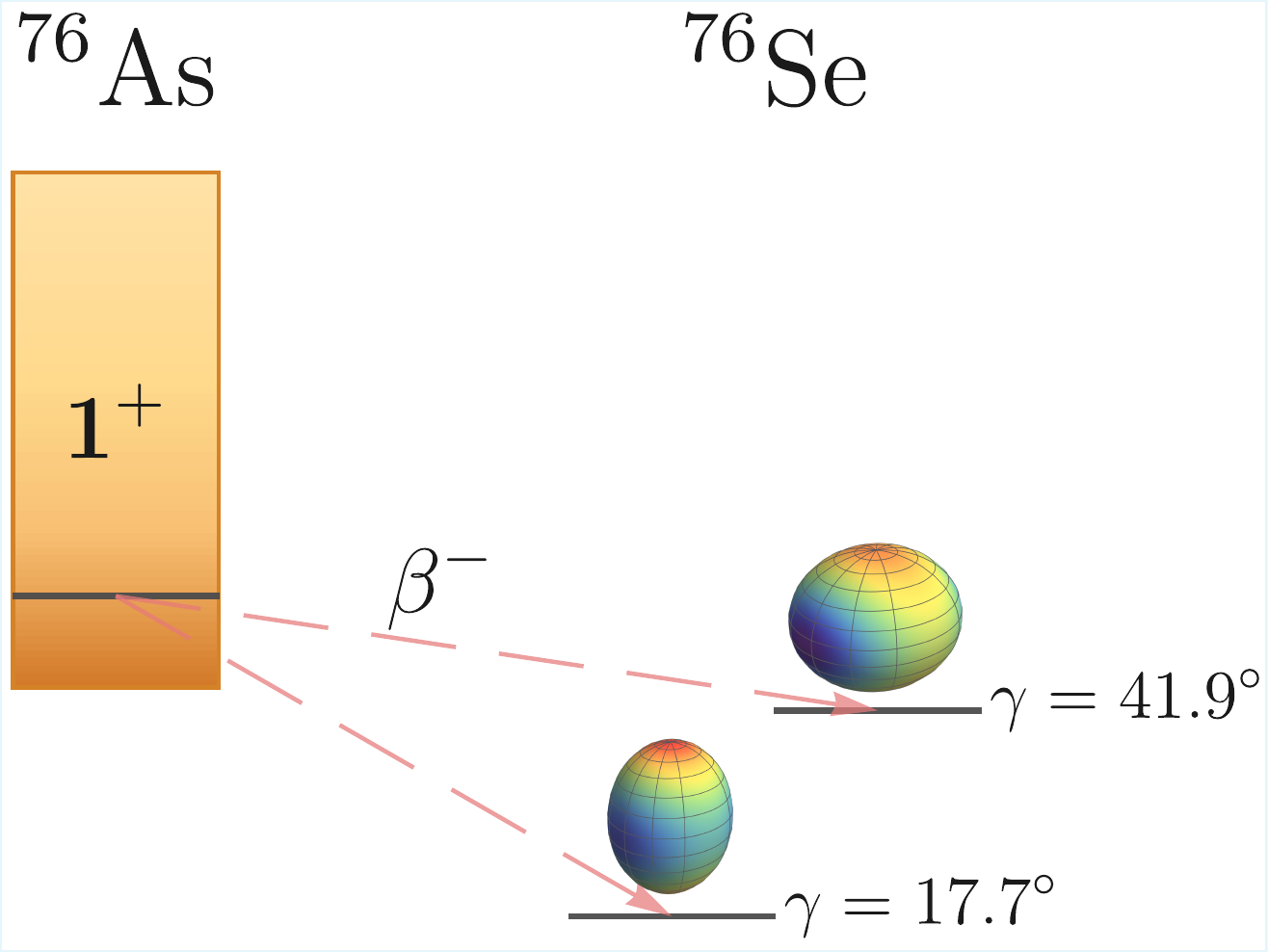}
\caption{(Color online) The lowest mean-field configuration of $\se$ in the $[6,0]$ occupation exhibits clear shape coexistence. Within the C60 virtual decay path, the two coexisting minima act as distinct final state candidates and give rise to different values of the $\db$ nuclear matrix element.}
\label{fig:Fig06}
\end{figure}

In the C60 scenario, a considerably more complex picture emerges. The analysis of the $\se$ mean-field $[6,0]$ configuration space 
representing the ground-state revealed the classic example of shape coexistence at low energies.
According to our calculations there are two configurations having the same parity-signature configuration being very close 
in energy but differing  significantly in the triaxiality parameter $\gamma$, as presented in the Table \ref{tab:76Se_shape}.
\begin{table}[h!]
\caption{\label{tab:76Se_shape} Energy and deformation parameters of 
the coexisting minima 
for the lowest (ground-state) mean-field $[6,0]$ configuration in $\se$, calculated using the SV force.}
\begin{ruledtabular}
\begin{tabular}{lccc}
$[\nu,\pi]$ & $E$ (MeV) & $\beta$ & $\gamma$ (deg) \\
\hline
$[6,0]$ & -643.79  & 0.30 & 41.9 \\
$[6,0]$ & -644.03 & 0.29 & 17.7 \\
\end{tabular}
\end{ruledtabular}
\end{table}
As shown in the table, they differ in energy by approximately 200\,keV what is very well inside typical uncertainties of any mean-field
calculations. Even a small change of the parameters of the Skyrme force can, in principle, change the prolate-oblate balance and invert the
minima.  Moreover, due to the shape difference and, most likely,  to too weak a pairing in our model, mixing between these states is extremely 
weak. Consequently, one cannot exclude that each of them may serve as a daughter state for $\as \rightarrow \se$ virtual 
transitions, as depicted in the Fig.~\ref{fig:Fig06}. Hence, in the following we shall consider two possible 
candidates for the $[6,0]$ ground-state of $\se$, the less and more triaxial. 

\begin{figure}
    \centering
    \includegraphics[width=0.35\textwidth]{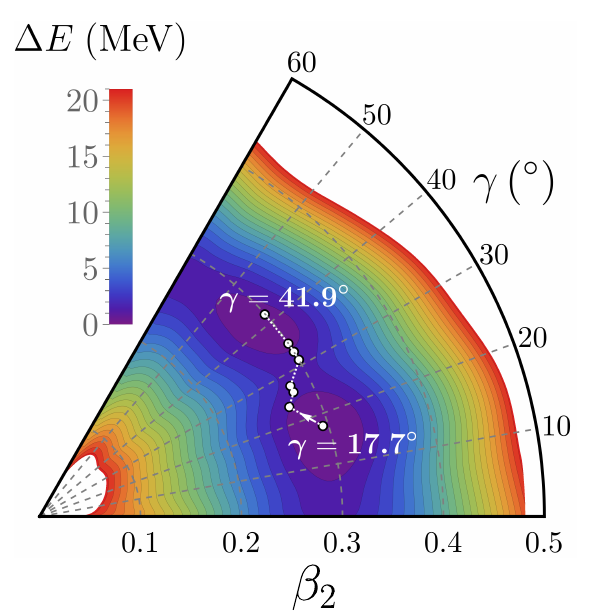}
    \caption{(Color online) Nuclear potential energy surface in the Bohr $(\beta_2,\gamma)$ deformation parameters for the $\se$ ground state $[6,0]$ configuration. The intermediate points denote states obtained with an axial quadrupole constraint. Due to the weak mixing between $\gamma = 17.7^{\circ}$ and $\gamma=41.9^{\circ}$ minima, they constitute two separate ground state candidates. The excitation energy has been renormalized to the HF minimum of $\gamma=17.7^{\circ}$.}
    \label{fig:Fig07}
\end{figure}

In the course of the study, it was found that shape coexistence for the $[6,0]$ configurations also frequently appears in excited 
configurations. To investigate the effects of shape coexistence, shape mixing, and shape vibrations on the calculated matrix element, 
the configuration space for the C60 path was defined slightly differently. It includes 49 configurations which, for the sake of further 
discussion, are divided into the following groups:

\begin{itemize}
\item
{\bf Group 1} includes the two coexisting ground states and thirty excited seniority-zero in-shell $nn$-, $pp$-, and $np$-pairing 
excitations. Seven configurations exhibit shape coexistence.
\item
{\bf Group 2} comprises nine configurations corresponding to the lowest seniority-zero $2p$–$2h$ $nn$-pairing excitations
$(\nu g_{9/2})^2 \rightarrow (\nu (sdg))^2$ across the $N = 50$ gap, as well as the lowest $nn$-pairing excitation
$(\nu f_{7/2})^2 \rightarrow (\nu (pf))^2$ across the $N = 28$ gap.
\item
{\bf Group 3} contains six constrained Hartree–Fock (HF) solutions along the barrier separating the coexisting ground-state minima, as shown in Fig.~\ref{fig:Fig07}.
\end{itemize}

Group 3 was included specifically to investigate the possibility of shape mixing between the coexisting ground-state minima through the quadrupole field. The corresponding configurations are shown on the potential energy surface in Fig.~\ref{fig:Fig07}. They were calculated using constrained Hartree–Fock theory with an axial quadrupole constraint, $\mathcal{C}(\hat{Q}_{20} - \langle \hat{Q}_{20} \rangle)$, where $\mathcal{C}$ and $\langle \hat{Q}_{20} \rangle$ are externally provided stiffness parameter and target quadrupole moment, respectively.

\section{Nuclear Matrix Elements and Decay Path Dependence}{\label{results}}

In this section, we present the results for the nuclear matrix element of the $\ger \rightarrow \se$ $\db$ decay. The results are given for two separate scenarios, depending on whether the decay proceeds along the C42 or C60 path. The theoretical uncertainties of the underlying Skyrme interaction require us to treat these scenarios as equally plausible, despite the fact that the SV Skyrme interaction shows a slight preference for the C42 path.

\subsection{Scenario C42}

\begin{figure}[h!]
\vspace{0.2cm}
\centering
\includegraphics[width = 0.44\textwidth]{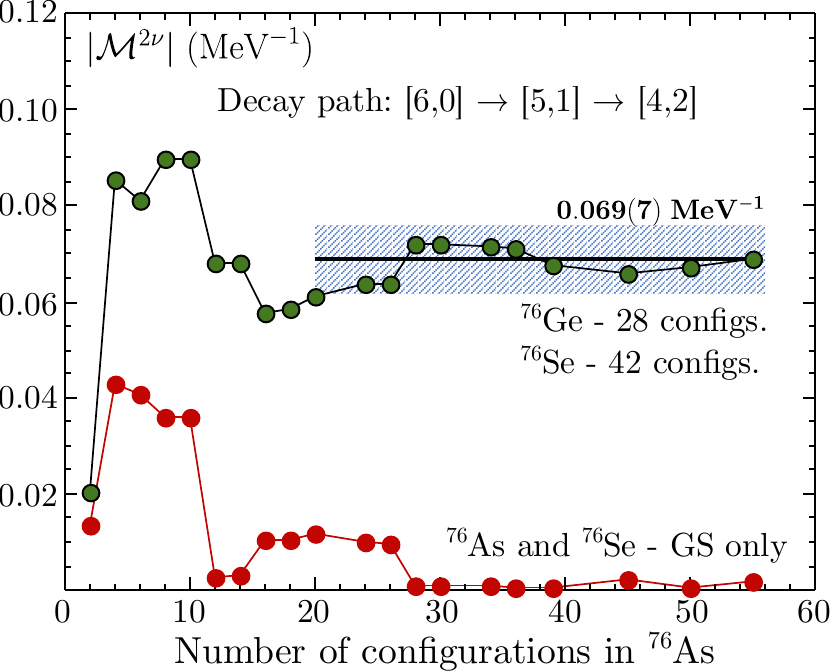}
\caption{\label{fig:Fig08} (Color online)
Stability of the $|\nm|$ matrix element as a function of the number $n$ of configurations in the virtual nucleus $^{76}$As. The plot shows two scenarios: no mixing in either the parent or daughter nuclei (red dots), and full correlations in both (black dots). The configurations in $^{76}$As are ordered in ascending HF excitation energy.
}
\end{figure}

Fig.~\ref{fig:Fig08} illustrates the stability of our calculations with respect to the number of virtual $1^+$ excitations in $^{76}$As. To highlight the role of configuration mixing, the plot contains two curves. One corresponds to the case without mixing in either the parent or daughter nuclei, while the other represents fully correlated parent and daughter nuclei.

As is clearly visible, the fully correlated calculations are well converged for $n \geq 30$, yielding $|\nm| = 0.069(7),\text{MeV}^{-1}$. The uncertainty, of order $10\%$, is conservatively assumed to account for possible contributions from missing or not fully converged configurations. This assumed uncertainty is significantly larger than the standard deviation calculated for $n \geq 30$ (see Fig.~\ref{fig:Fig06}), $\sigma_{n \geq 30} = 0.002\,\text{MeV}^{-1}$.

\begin{figure}[h!]
\vspace{0.2cm}
\centering
\includegraphics[width = 0.48\textwidth]{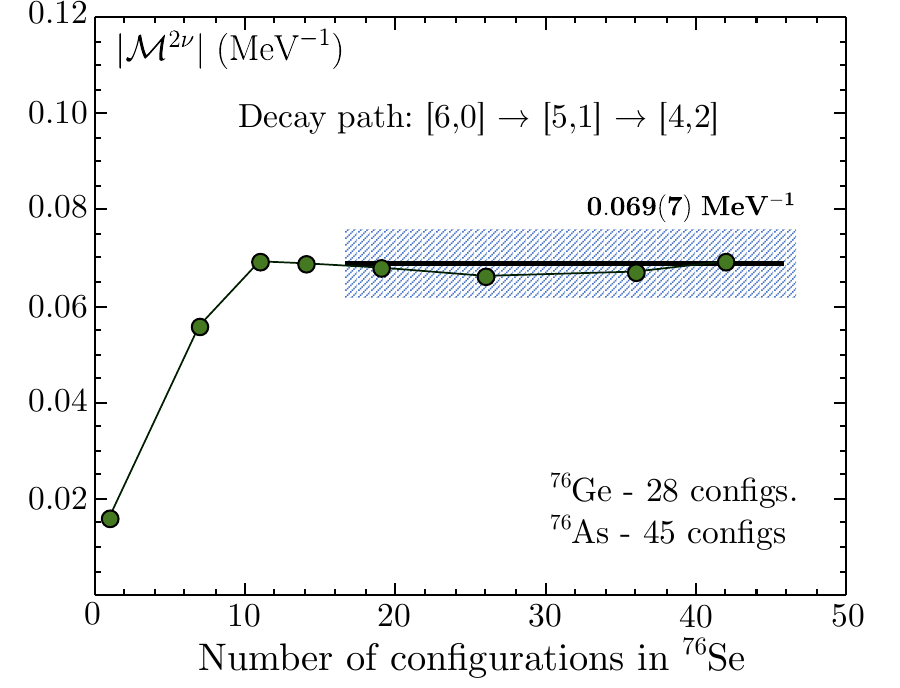}
\caption{\label{fig:Fig09} (Color online)
Stability of the $|\nm|$ matrix element as a function of the number $n$ of configurations in the daughter nucleus $^{76}$Se.
The configurations in $^{76}$Se are grouped by type and, within each group, ordered by increasing HF excitation energy.
}
\end{figure}

Fig.~\ref{fig:Fig08} also demonstrates that configuration mixing plays a crucial role in both the parent and daughter nuclei. To further elucidate the role of mixing in the parent and daughter systems, we present in Fig.~\ref{fig:Fig09} the stability of our calculations with respect to the number of excitations in the daughter nucleus $^{76}$Se. In these calculations, both the parent nucleus and the virtual nucleus are fully correlated.

The configurations in $^{76}$Se are ordered differently compared to the case discussed above; they are grouped by type. This grouping highlights the impact of different classes of configurations on the calculated matrix element. In particular, we observe that in-shell $nn/pp$ pairing correlations - i.e., configurations belonging to Groups 1 and 2 - essentially saturate the matrix element. All other in-shell configurations, including $np$ pairing as well as cross-shell $nn/pp$ pairing excitations (Groups 3–7), affect the value of $|\nm|$ only marginally and can, in principle, be neglected.

\begin{figure}[h!]
\vspace{0.2cm}
\centering
\includegraphics[width = 0.48\textwidth]{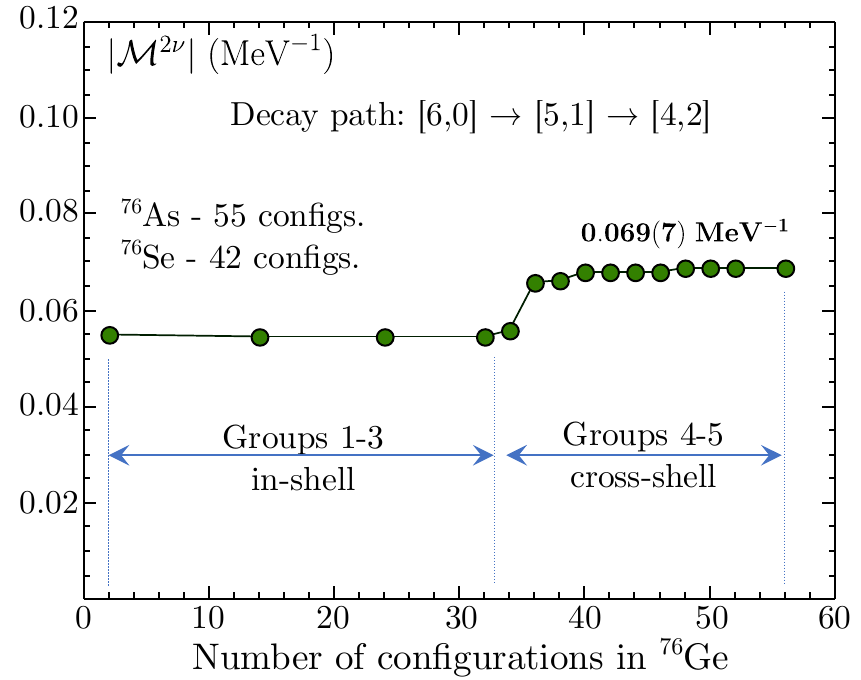}
\caption{\label{fig:Fig10} (Color online)
Stability of the $|\nm|$ matrix element as a function of the number $n$ of configurations in the parent nucleus $^{76}$Ge.
The configurations in $^{76}$Ge are grouped by type and, within each group, ordered by increasing HF excitation energy.
}
\end{figure}

Finally, Fig.~\ref{fig:Fig10} illustrates the stability of our calculations with respect to the number of excitations in the parent nucleus $^{76}$Ge. In these calculations, both $^{76}$As and $^{76}$Se are fully correlated. The configurations in $^{76}$Ge are grouped by type.

As shown in the figure, the in-shell excitations belonging to Groups 1–6 do not play a significant role. The matrix element is dominated by the ground-state configuration (yielding $|\nm| \approx 0.055,\text{MeV}^{-1}$) and by cross-shell excitations from Group 7. Notably, a single configuration from Group 7 - although not the lowest in energy within our model - is sufficient to increase the matrix element to $|\nm| \approx 0.066,\text{MeV}^{-1}$, i.e., very close to its final value.

\subsection{Scenario C60}

As already discussed, the lowest $[6,0]$ configuration exhibits shape coexistence. The two coexisting ground-state-like configurations -- i.e., configurations having the same occupancies within each parity-signature block -- differ by approximately 200\,keV and correspond to $\gamma = 17.7^{\circ}$ (global minimum) and $\gamma = 41.9^{\circ}$ (excited local minimum), respectively. 

\begin{figure}[h]
\vspace{0.2cm}
\centering
\includegraphics[width = 0.48\textwidth]{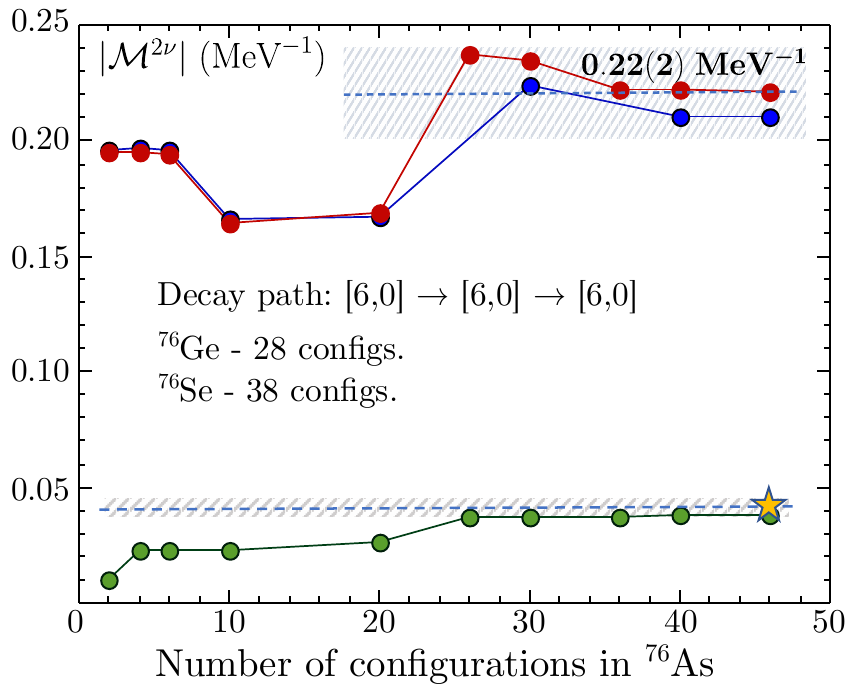}
\caption{\label{fig:Fig11} (Color online) Stability of the $| \nm |$ matrix element with respect to the number of configurations in 
the virtual nucleus $^{76}$As for the C60 path. Green dots correspond to the scenario involving the global minimum at 
$\gamma = 17.7^{\circ}$ and correlated parent and daughter nuclei. The yellow star marks the result involving  
six constrained HF solutions (Group 3) in the configuration space of  $^{76}$Se. The curves marked by black (red) dots correspond 
to decay to the more triaxial minimum at $\gamma = 41.9^{\circ}$ and were  computed using uncorrelated (correlated) parent and 
daughter nuclei, respectively.  The configurations in the virtual nucleus are ordered in ascending order according to their mean-field energies.  
}
\end{figure}

The stability of the $\nm$ corresponding to the scenario involving the global minimum at $\gamma = 17.7^{\circ}$ (green dots) is illustrated in Fig.~\ref{fig:Fig11} as a function of the number of configurations in $^{76}$As. As seen in the figure, $|\nm|$ reaches a plateau at approximately 0.038(4),MeV$^{-1}$ already for $n \approx 20$ configurations. Increasing the configuration space of the daughter nucleus by adding six additional constrained HF solutions from Group 3 increases the NME to 0.040(4)\,MeV$^{-1}$ (the point marked by a yellow star), i.e., it has a relatively small impact on the calculated value. This result suggests that shape vibrations have probably rather modest impact on the calculated NMEs,
though the point requires further studies. 

The two other curves, marked by black and red dots, respectively, correspond to decay to the more triaxial minimum at $\gamma = 41.9^{\circ}$. The results represented by black (red) dots correspond to uncorrelated (correlated) parent and daughter nuclei, respectively. It is striking that the calculated $\nm$ is large, of the order of 0.22(2)~MeV$^{-1}$. Moreover, unlike in all other cases analyzed here, the result appears to be fairly independent of correlations in the ground states of the parent and daughter nuclei. The case is evidently dominated by a single state in each
of the participating nuclei.   

It is worth emphasizing that, despite the overall large configuration space, it remains significantly smaller than those used in shell-model-based frameworks. Furthermore, excluding the configuration groups in the parent and daughter nuclei that mix only weakly with the dominant configurations allows the effective configuration space to be reduced even further.

\section{Summary and conclusions}\label{summary}

We present a theoretical study of the two-neutrino $0^+ \rightarrow 0^+$ double beta decay of $^{76}$Ge within the No-Core Configuration-Interaction (DFT-NCCI) framework based on the Skyrme SV density functional. Our calculations reveal that the configuration space splits to a high
precision into blocks characterised by the $[n,m] \equiv [(\nu g_{9/2})^n, (\pi g_{9/2})^m]$ occupancy of the $0g_{9/2}$ intruder orbital in
the daughter nucleus.

Taking into account the typical uncertainties of the underlying Skyrme interaction, we identify three different decay scenarios. 
In this work, we focus on the two decay paths that both originate from the $[6,0]$ ground state in $^{76}$Ge, the 
configuration favored by both the SV and SLy4 functionals.

The energetically favored scenario corresponds to the $[4,2]$ occupation (the C42 scenario) in $\se$. For this scenario, 
we obtain $|\mathcal{M}^{2\nu}| = 0.069(7)$~MeV$^{-1}$, which is consistent with other calculations based on energy-density-functional frameworks. However, this value is considerably smaller than the empirical estimate reported by A. S. Barabash~\cite{(Bar20)}, namely $|\mathcal{M}^{2\nu}| = 0.204(14)$~MeV$^{-1}$; see Tab.~\ref{tab:M2nu_76Ge_comparison} and Fig.~\ref{fig:Fig13}.

\begin{figure}[h!]
\centering

\subfloat[C42 scenario.]{
	\label{C42-arrow}
	\includegraphics[width=0.46\textwidth]{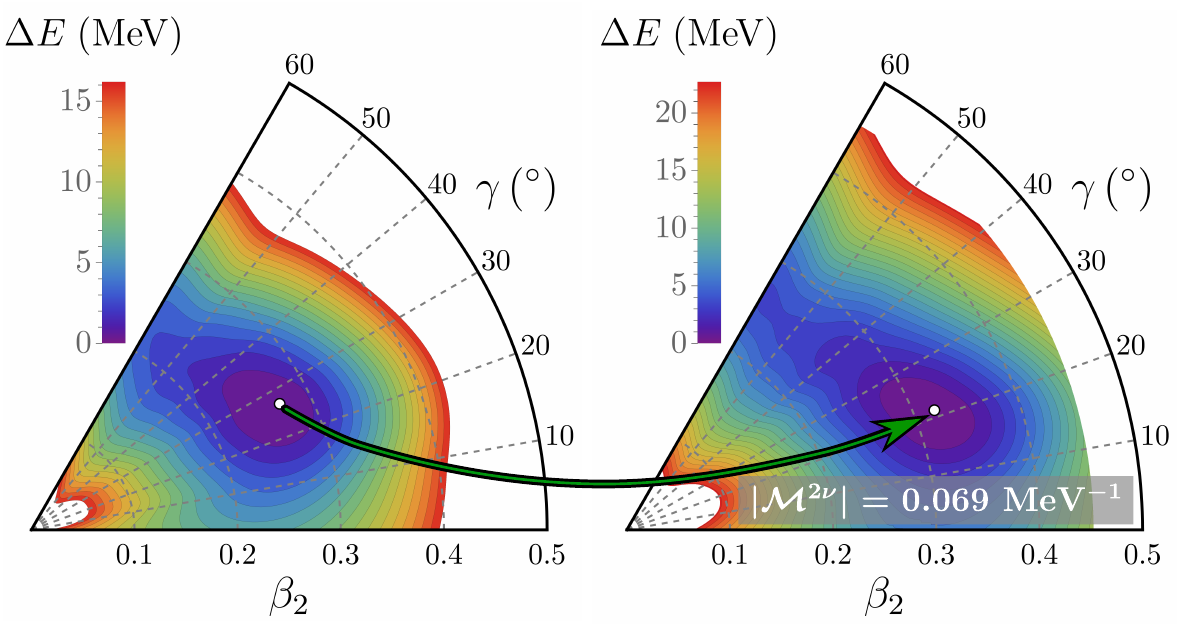} } 

\subfloat[C60 scenario.]{
	\label{C60-arorw}
	\includegraphics[width=0.47\textwidth]{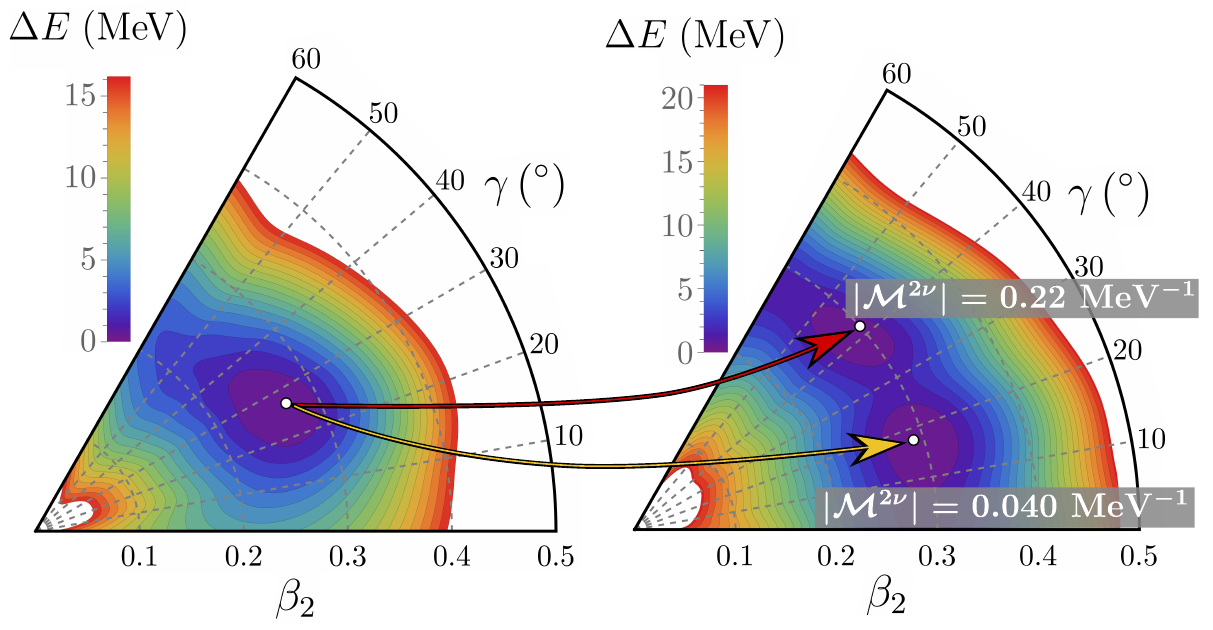} } 
\caption{(Color online) Results for the $2\nu\beta\beta$ decay in the C42 (a) and C60 (b) scenarios summarized on the corresponding nuclear potential energy surfaces. The green (orange, red) arrow denotes the most (less, least) energetically favored decay path according to the SV Skyrme interaction.}
\label{fig:Fig12}
\end{figure}

\begin{figure}[htb]
%\vspace{0.1cm}
\centering
\includegraphics[width = \columnwidth]{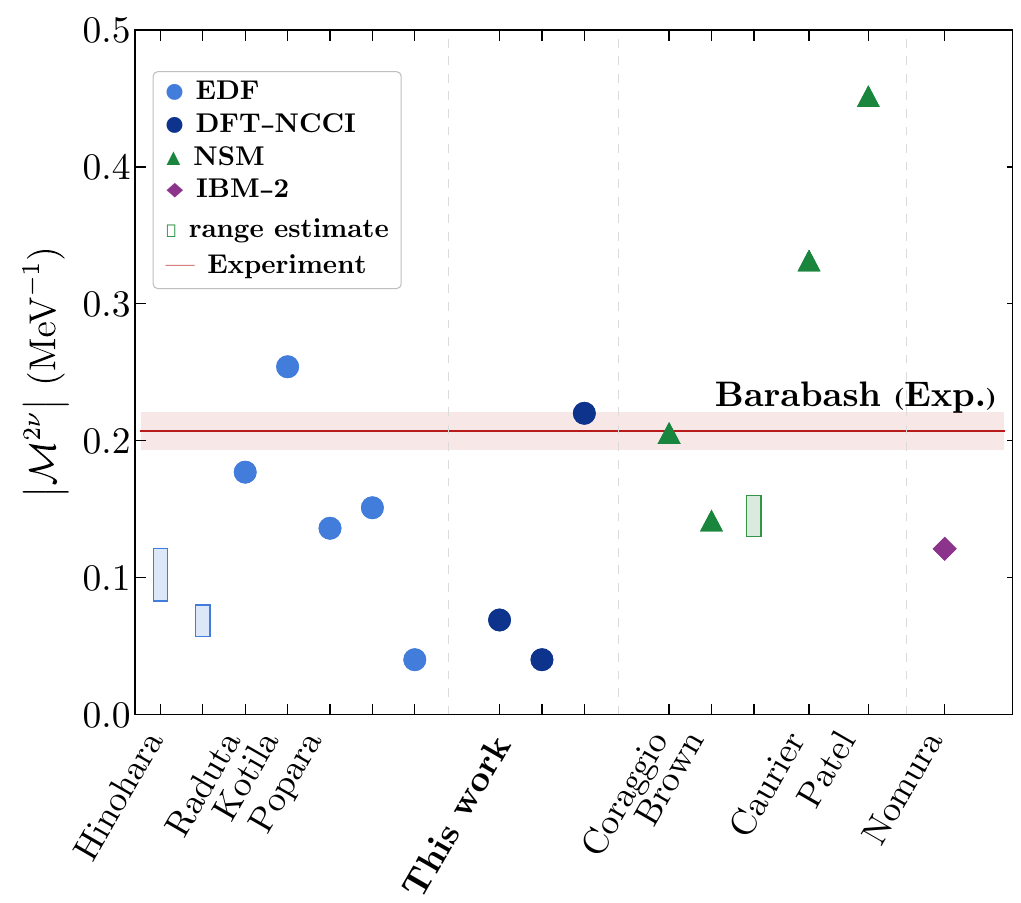}
\caption{\label{fig:Fig13} (Color online) A summary of selected NME calculated using different models. All values are renormalized to 
the effective axial-vector coupling constant $g^{\text{(eff)}}_{\text{A}} =1$ to emphasize the nuclear-structure content. The empirically derived effective NME from 
Ref.~\cite{(Bar20)} is indicated by a horizontal red line, with its uncertainty shown as a shaded red area. The markers, from left to right, represent: EDF-based results from Ref.\cite{(Hin22),(Rad13),(Kot12),(Pop22)}; results obtained in this work for C42, C60 ($\gamma=17.7^\circ$, $\gamma=41.9^\circ$) scenarios respectively; shell-model results from Refs.~\cite{(Cor19),(Bro15a),(Kos22),(Pat24)}; and results obtained using IBM-2 framework reported 
in Refs.~\cite{(Nom24)}. For clarity, only the first marker in each group is labeled by the first author’s name.
}
\end{figure}

The alternative scenario corresponds to $[6,0]$ occupation (C60 scenario). In this case our calculations reveal two coexisting ground-state
configurations  having the same occupations in terms of parity-signature quantum numbers but differing in triaxiality. 
These mean-field solutions differ by approximately 200\,keV and correspond to $\gamma=17.7^{\circ}$ (global minimum) and     
$\gamma = 41.9^{\circ}$ (the lowest local minimum), respectively, as presented on the Fig.~\ref{fig:Fig12}. The resulting $2\nu\beta\beta$ nuclear matrix element is found to depend strongly on the triaxiality, yielding $|\mathcal{M}^{2\nu}| = 0.040(4)$~MeV$^{-1}$ at $\gamma = 17.7^\circ$ and $|\mathcal{M}^{2\nu}| = 0.22(2)$~MeV$^{-1}$ at $\gamma = 41.9^\circ$. The first value is  consistent with other calculations based on energy-density-functional 
frameworks but is considerably smaller than the empirical estimate while the second result matches quite well the 
experimental  estimate  $0.204(14)$~MeV$^{-1}$ quoted above, see Tab.~\ref{tab:M2nu_76Ge_comparison} and Fig.~\ref{fig:Fig13}. 
All calculated NMEs quoted above are renormalized to $g_A^{\text{(eff)}}=1$.

Our results are subject to theoretical uncertainties, which we estimate to be no greater than 10\%. 
In our opinion, this represents a rather conservative estimate. The calculations appear to be well converged; 
but  we cannot rule out the possibility that some relevant configurations are missing from the 
model space, which could affect the results. Another potential source of uncertainty in the normalization of the spectrum 
of $|I=1^+\rangle $ states in the virtual intermediate nucleus which was assumed 100\,keV. 
The calculations show that it has a moderate impact on the calculated NME. For example, in the C42 scenario, changing the 
normalization from $\Delta E (1^+) = 0.100$,MeV to 0.050 (0.300)\,MeV changes $|\nm|$ from 0.069\,MeV$^{-1}$ 
to 0.070 (0.065)\,MeV$^{-1}$, respectively -- i.e., well within the assumed theoretical uncertainties. An analogous change in 
the normalization for the C60 scenarios leads to similar conclusions -- the changes remain well within the assumed uncertainties.

\begin{table}[htbp]
\caption{$\nm$ estimation for $^{76}$Ge decay within various nuclear models, renormalized to $g_{\rm A}^{\rm{(eff)}}=1$. See the references for dimensionless $\mathcal{M}^{2\nu}$ values.}
\label{tab:M2nu_76Ge_comparison}
\begin{ruledtabular}
\small
\setlength{\tabcolsep}{4pt}
\begin{tabular}{l@{\hskip 6pt}l@{\hskip 6pt}l}
\textbf{Reference} & \textbf{Method} & $\boldsymbol{|\nm|}$ (MeV$^{-1}$) \\
\hline
Barabash \cite{(Bar20)}             & Experiment  & $0.204 \pm 0.014$ \\
Hinohara \& Engel \cite{(Hin22)}        & QTDA   & $0.083$--$0.121$ \\
                               & QRPA   & $0.057$--$0.080$ \\
Raduta et al. \cite{(Rad13)}  & RPA    & $0.177$ \\
Kotila et al. \cite{(Kot09)} & pnMAVA & $0.254 \pm 0.01$ \\
Popara et al. \cite{(Pop22)} & QRPA   & $0.136$ \\
                               & pnQRPA & $0.151$ \\
                               & DD-ME2 & $0.04$ \\
Coraggio et al. \cite{(Cor19)} & ISM & 0.204 \\
Brown et al. \cite{(Bro15a)} & CI & $0.140 \pm 0.005$ \\
                               & QRPA & $0.13$--$0.16$ \\
Caurier et al. \cite{(Cau12a)}* & NSM & 0.33 \\   
Patel et al. \cite{(Pat24)} & NSM & 0.45 \\  
Barea et al. \cite{(Bar15b)} & IBM-2 & 0.86 \\
Nomura \cite{(Nom22)} & IBM-2 & 0.121 \\          
\textbf{This work} & \textbf{DFT-NCCI} & $\boldsymbol{0.069(7)}$ (C42) \\
                               &        & $\boldsymbol{0.040(4)}$ (C60) \\
                               &        & $\boldsymbol{0.22(2)}$ (C60) \\
\end{tabular}
\end{ruledtabular}
\end{table}
\vspace{-0.4cm}
\begin{minipage}{0.9\textwidth}
\footnotesize
* repeated by Kostensalo et al \cite{(Kos22)}.
\end{minipage}
\vspace{0.1cm}

Table~\ref{tab:M2nu_76Ge_comparison} presents a comparison between our work and the values provided by other models, when renormalized to $q g_{\rm A} = 1$. This comparison illustrates that the $\ger \rightarrow \se$ decay remains theoretically difficult to describe, as the results are far more divergent than in the case of $^{48}\text{Ca} \rightarrow ^{48}\text{Ti}$ \cite{(Mis25b)}. In particular, the EDF-based results \cite{(Hin22),(Rad13),(Pop22)} depend strongly on the functional used. This is most evident in the study by Hinohara and Engel, where 10 different Skyrme parametrizations were employed, each predicting a different value of $\nm$.

On the other hand, Shell Model~\cite{(Bro15a),(Kos22),(Pat24)} and IBM-2 \cite{(Bar15b),(Nom22)} studies suggest a significantly different picture, in which the models tend to overestimate $\nm$ by a factor of $\approx 1.66$--$4$. A notable exception is the estimate provided by Coraggio 
{\it et al.}~\cite{(Cor19)} within the Interacting Shell Model framework (ISM), which predicts $|\nm| = 0.196 \pm 0.007~\mathrm{MeV}^{-1}$, in agreement with the experimental reference reported by Barabash~\cite{(Bar20)}.

A plausible underlying reason for this discrepancy is the rich nuclear structure of the $\ger - \se$ system. In particular, the dominant triaxial character of these nuclei and the shape-coexistence phenomena result in complex mixing of single-particle orbitals. This situation is qualitatively  different from that in the (predominantly) axially deformed $^{48}\text{Ca} -  ^{48}\text{Ti}$ system, where the shell structure is dominated by the $0f_{7/2}$ and $pf$ orbitals above the $N(Z) = 28$ shell gap -- features that are well captured by most theoretical approaches.
This explains consistency of theoretical predictions for the  $^{48}\text{Ca} \rightarrow  ^{48}\text{Ti}$ decay.

As already mentioned in Sec.~\ref{intro}, the theoretical interest in $\db$ decays comes from the potential connection between the corresponding nuclear matrix elements for $2\nu\beta\beta$ and $0\nu\beta\beta$. In particular, two structural effects are believed to play a key role in determining the magnitude of $\zb$ -- nuclear deformation and pairing correlations.

Analysis of the effective quadrupole interaction $\hat{V}_{QQ}=~\chi~\hat{Q}\cdot\hat{Q}$ and of the resulting deformation difference $\Delta\beta$ between the parent and daughter nuclei, as in \cite{(Men11a)}, shows that $\zm$ is strongly suppressed with increasing $\Delta\beta$. This correlation has also been observed in $\db$-active nuclei \cite{(Sim04)}. Interestingly, our results seem to support recent studies \cite{(Jia17),(Wan24a)}, which suggest that the triaxiality parameter $\gamma$ has a nontrivial effect on the value of $\nm$. This point deserves further investigation; density-functional-based frameworks, such as DFT-NCCI, are particularly well suited to address this issue.

In summary, our calculations reveal significant challenges in the precise determination of $|\mathcal{M}^{2\nu}|$ for the $^{76}$Ge decay, offering an explanation for the difficulties in theoretical modeling of this quantity, which varies by almost an order of magnitude in the available
literature. The structural complexity, triaxiality, and shape coexistence identified in the analyzed nuclei imply a strong sensitivity to fine details of the interaction and configuration mixing.

One factor that is likely underestimated in our model, and that may enhance mixing between the different decay paths discussed here, is pairing. Pairing correlations enter our DFT-NCCI calculations through configuration mixing governed by the SV parametrization of the Skyrme interaction, which is characterized by a very low effective mass. This, in turn, lowers the density of states and suppresses nuclear pairing.

The construction of a projected Skyrme-based framework that incorporates realistic nuclear superfluid properties would provide a major advantage for future $\beta\beta$ studies. Pairing correlations are predicted to strongly enhance the calculated values of $\beta\beta$ matrix elements, as discussed extensively in Refs.~\cite{(Cau08),(Men11)}. This challenge remains ahead of us, as does the estimation of the effective axial coupling strength, $g_{\rm A}^{\rm (eff)} = q g_{\rm A}$, where $q$ is the in-medium quenching factor for the axial current.

%\begin{acknowledgments}

%\end{acknowledgments}
\bibliographystyle{apsrev4-2}
\bibliography{apssamp,TEDMED,jacwit34}% Produces the bibliography via BibTeX.

\end{document}